\documentclass{article}
\usepackage{epsfig}
\begin{document}

\title{HST Observations of Black Hole X-Ray Transient Outbursts}
\author{Robert I. Hynes$^1$, Carole A. Haswell$^2$}
\date{}

\maketitle

$^1$Department of Physics and Astronomy, University of
Southampton, Southampton, SO17 1BJ, UK.\\
$^2$Department of Physics and Astronomy, The Open
University, Walton Hall, Milton Keynes, MK7 6AA, UK.
\vspace*{-4mm}
\section{Introduction}
\vspace*{-3mm}

Black hole X-ray transients (BHXRT; also known as Soft X-Ray
Transients and X-Ray Novae; see Tanaka \& Shibazaki 1996) contain a
compact object accreting from a low-mass star via Roche lobe overflow
and an accretion disk.  They differ from other X-ray binaries in that
significant X-ray emission is seen only in well defined outbursts
lasting up to two years and recurring on timescales of decades.  The
currently favored model for these outbursts is a variant of the disk
instability model originally developed for dwarf nova outbursts
(Cannizzo 1999).  In $>70$\,\% of cases, the compact object is
dynamically measured to be a black hole (Charles 1998).  These objects
are thus the best place to study stellar mass black holes.  In
particular, there has recently been much theoretical work on advective
accretion flows, and variations on this theme, in which the black hole
`silently' accretes material through its event horizon with little
emission (Narayan, Mahadevan \& Quataert 1998).  The reality of such
flows remains controversial.

The first {\it HST} observations of a BHXRT were of X-ray Nova Muscae
1991 through Director's Discretionary Time (Cheng et al.\ 1992).  
After this, a formal target-of-opportunity program was established to
study BHXRT outbursts, originally using FOS and now STIS.  The current
program contains 20 STIS orbits spread over 7 visits spanning the
first 150 days of outburst.  Our goals include: to monitor changes in
the 1100--10300\,\AA\ spectrum through the outburst; to obtain a
high-quality far-UV spectrum to study the typically rich emission line
spectrum; to use the near-UV spectrum to fit the 2175\,\AA\
interstellar absorption feature and determine the interstellar
reddening; and to obtain UV data in TIMETAG mode for comparison with
simultaneous RXTE X-ray light curves and echo mapping.  For initial
visits on a bright target we also use medium-resolution echelle modes
to obtain UV line spectrum with velocity resolution
$<10$\,km\,s$^{-1}$.


\vspace*{-4mm}
\section{GRO\,J0422+32}
\vspace*{-3mm}

The first target for the target-of-opportunity program, GRO\,J0422+32
was discovered in August 1992 by {\it CGRO}/BATSE.  The primary is
likely a 4--12\,M$_{\odot}$ black hole and the companion an M0--M4
dwarf in a 5.1\,hr low-inclination orbit.  The outburst lightcurve is
the best studied of any BHXRT.  GRO\,J0422+32 was observed by {\it
HST}/FOS in a single visit at the end of the outburst, in 1994.  A
full description of this observation is given by Hynes \& Haswell
(1999).

The outburst behavior, illustrated in Fig.\ 1, is complex, dividing
into two phases spanning two years.  The early plateau phase
($\sim200$\,d) showed a steady exponential decline.  The later phase
shows the same decay rate at lower level, with several superposed
mini-outbursts.  Two mini-outbursts were definitely detected,
separated by 120\,d.  The near-quiescent {\it HST} observations appear
to have detected a third mini-outburst, or a flat plateau at end of
outburst.  This occurs $\sim240$\,d after the previous mini-outburst,
consistent with twice the 120\,d recurrence time.

The dereddened spectral energy distribution (SED) peaks at blue
wavelengths.  It appears sharper peaked than a black body; a
self-absorbed synchrotron spectrum fits much better.  Advective
accretion flow models predict that optical emission is dominated by
synchrotron, with similar peak wavelength to that seen.  The {\it HST}
observation may therefore indicate that an advective flow has already
formed at the end of the outburst.

\begin{figure}
\begin{minipage}[t]{3.37in}
\vspace*{-2.36in}
\epsfig{angle=90,width=3.37in,file=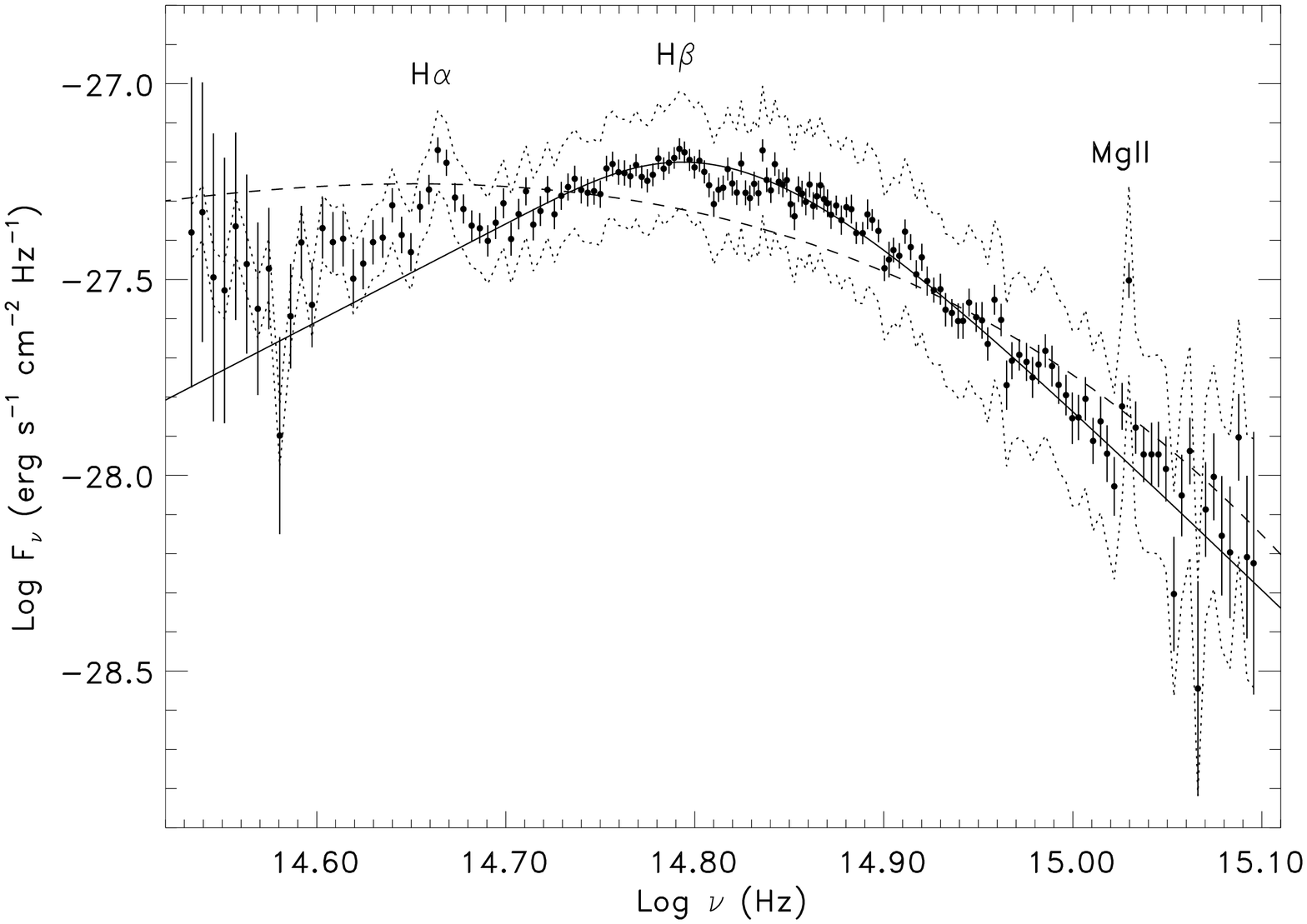}
\hspace*{-2.67in}
\begin{minipage}[t]{1.6in}
\vspace*{-1.5in}
\epsfig{angle=90,width=1.6in,file=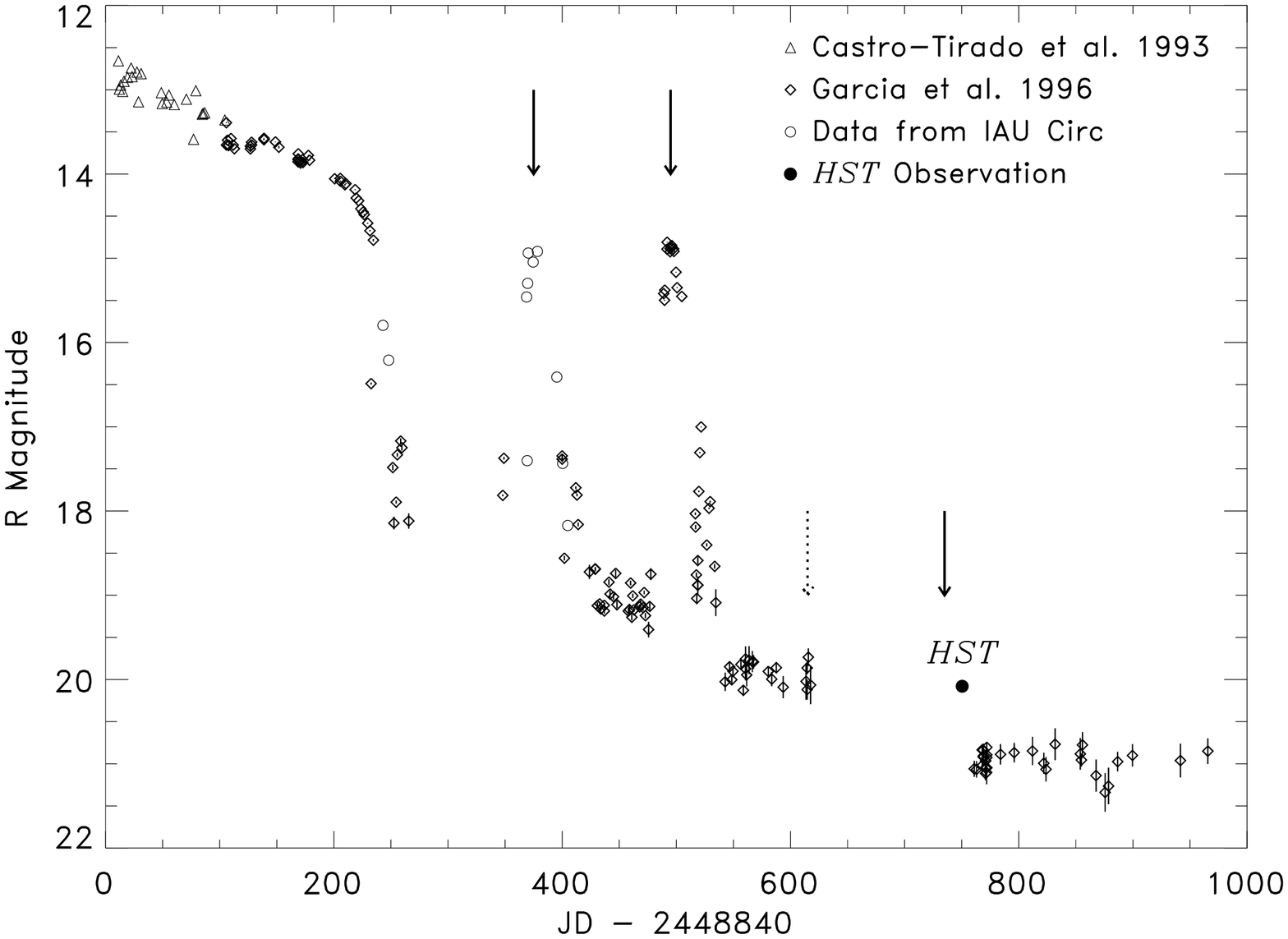}
\end{minipage}
\end{minipage}
\epsfig{angle=90,width=3.15in,file=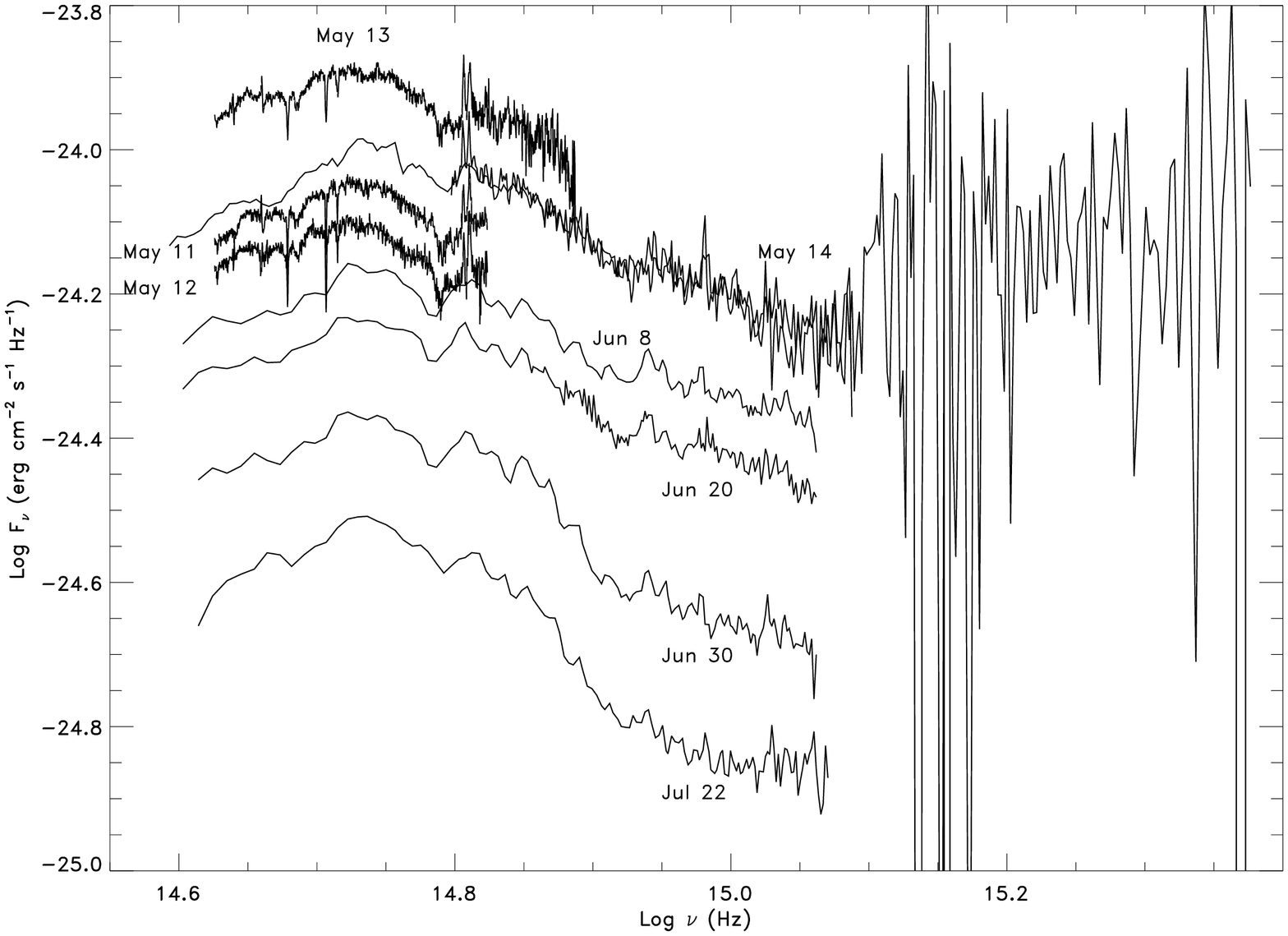}
\caption{Left: Dereddened {\it HST} spectral energy distribution (SED) of
GRO\,J0422+32 near quiescence, with the spectrum of the companion star
subtracted.  Models are a black body (dashed) and self-absorbed
synchrotron emission (solid).  Inset shows the outburst lightcurve,
and the timing of the {\it HST} observation.  Right: Dereddened SEDs of
GRO\,J1655--40 at several epochs.  Spectra from May 11--13 are from
the AAT, others are {\it HST}.}
\end{figure}

\vspace*{-4mm}
\section{GRO\,J1655--40}
\vspace*{-3mm}

This object was discovered by {\it CGRO}/BATSE in July 1994.  The
system is believed to comprise a 5--7\,M$_{\odot}$ black hole and an
F-type subgiant companion in a 2.6\,d orbit.  Several outbursts
occurred until 1997, during the first of which relativistic jets were
ejected.  The outburst optical and X-ray fluxes are not well
correlated.  In 1996, {\it HST} saw optical and UV fluxes drop whilst
X-rays rose (Hynes et al.\ 1998a).  This is difficult to reconcile
with simple outburst models with optical/UV emission produced by
reprocessing of X-ray emission via disk heating.  Echo mapping
indicates at least some reprocessing (see below); the most likely
solution is that reprocessing gets less effective later on.  A change
in the irradiation geometry, e.g.\ through warping of the disk (Esin,
Lasota \& Hynes 2000), could drive such a change.

The dereddened spectral energy distributions (Hynes et al.\ 1998a) are
very different to other systems (e.g.\ XTE\,J1859+226 shown below).
There is an optical component, peaking at $\log\nu = 14.6$--14.9 which
appears to be at least partly from the companion, since this is larger
and hotter than in other black hole systems.  The residual optical
spectrum looks like a relatively cool irradiated disk.  This makes
sense, as a large orbital period implies a large disk with an edge far
from the X-ray source.  The far-UV component looks more like the
$\nu^{1/3}$ spectrum of a viscously heated disk, but the transition
between optical and UV is sharper than expected.

Coordinated {\it HST} and {\it RXTE} observations revealed correlated
X-ray and optical/UV variability (Hynes et al.\ 1998b).  We were able
to cross-correlate the two light curves and derive optical lags of
10--20\,s, likely indicating light travel times within the system.
This is too short to originate on the companion star for which a
minimum lag of $\sim40$\,s is expected at the orbital phase observed.
Lags of 0--30\,s {\em are} expected from reprocessing in the disk.  It
thus appears that either the companion is shielded from X-rays, or for
some other reason does not reprocess variability.
\vspace*{-4mm}
\section{XTE\,J1859+226}
\vspace*{-3mm}

XTE\,J1859+226 was discovered in October 1999 by {\it RXTE}.  The
compact object is likely a black hole but neither this, nor the nature
of the companion have been confirmed.  An orbital period of 6.7\,hr
has been suggested (Uemura et al.\ 1999).  The fast rise, exponential
decay lightcurve is typical of BHXRTs, although the initial rise was
slower than normal.

Fitting the broad 2175\,\AA\ interstellar feature yields an
interstellar reddening of $E(B-V)=0.58\pm0.07$ (Hynes et al.\ 1999).
We used this value to deredden the spectra and derive the intrinsic
SEDs shown in Fig.\ 2.  SEDs from early in the outburst could be well
fitted by simple X-ray heated disk model, $T\propto R^{-3/7}$.  The
derived edge temperature remains constant as the system fades,
implying that the edge of the hot region moves inward as the X-rays
decline, likely either indicating a cooling wave or a self-shielding
effect.  The last SED is better fitted by a viscously heated disk
model with an edge temperature of $\sim8000$\,K, consistent with disk
instability models.

\begin{figure}
\epsfig{angle=90,width=3.25in,file=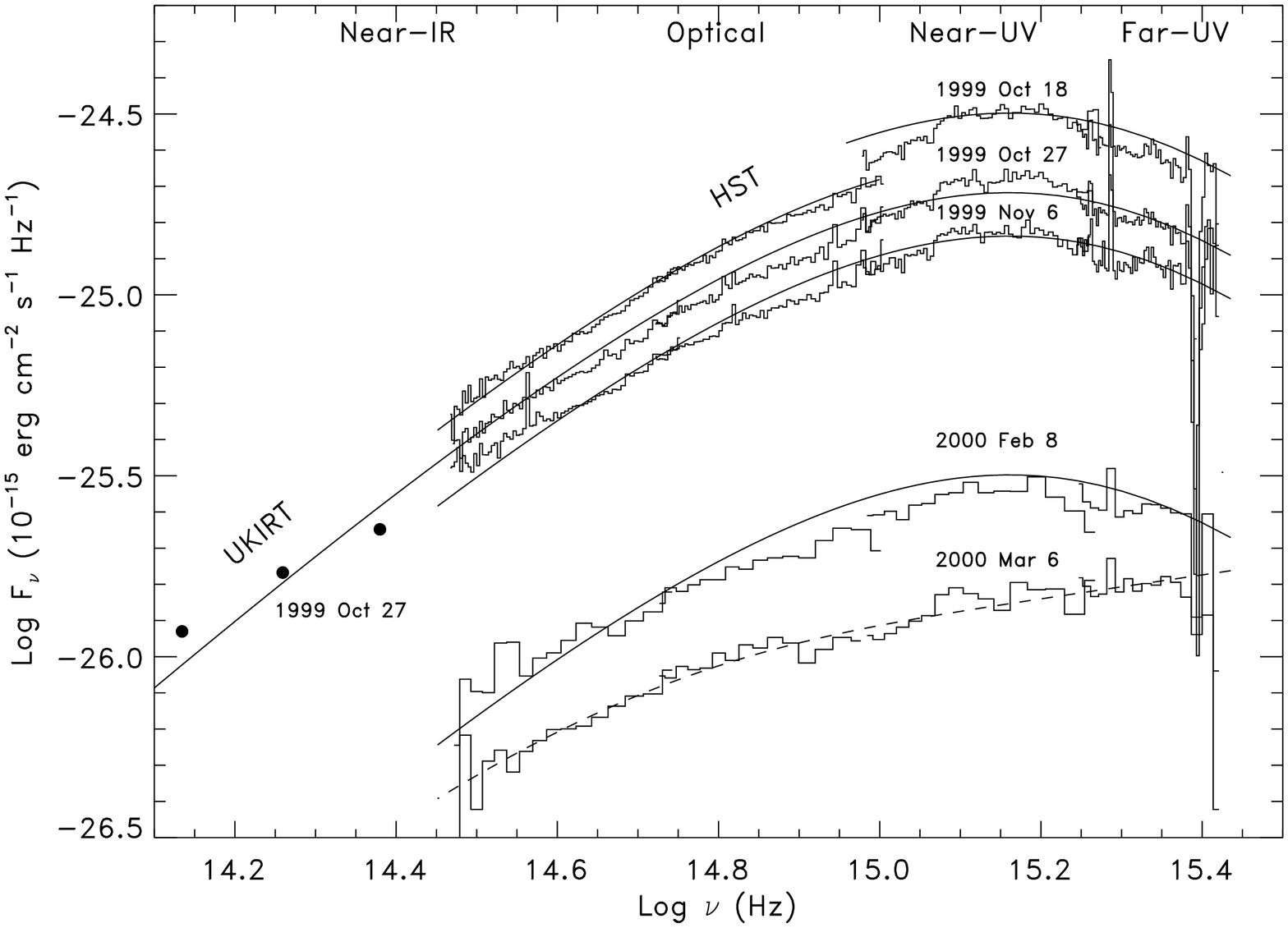}
\epsfig{angle=90,width=3.25in,file=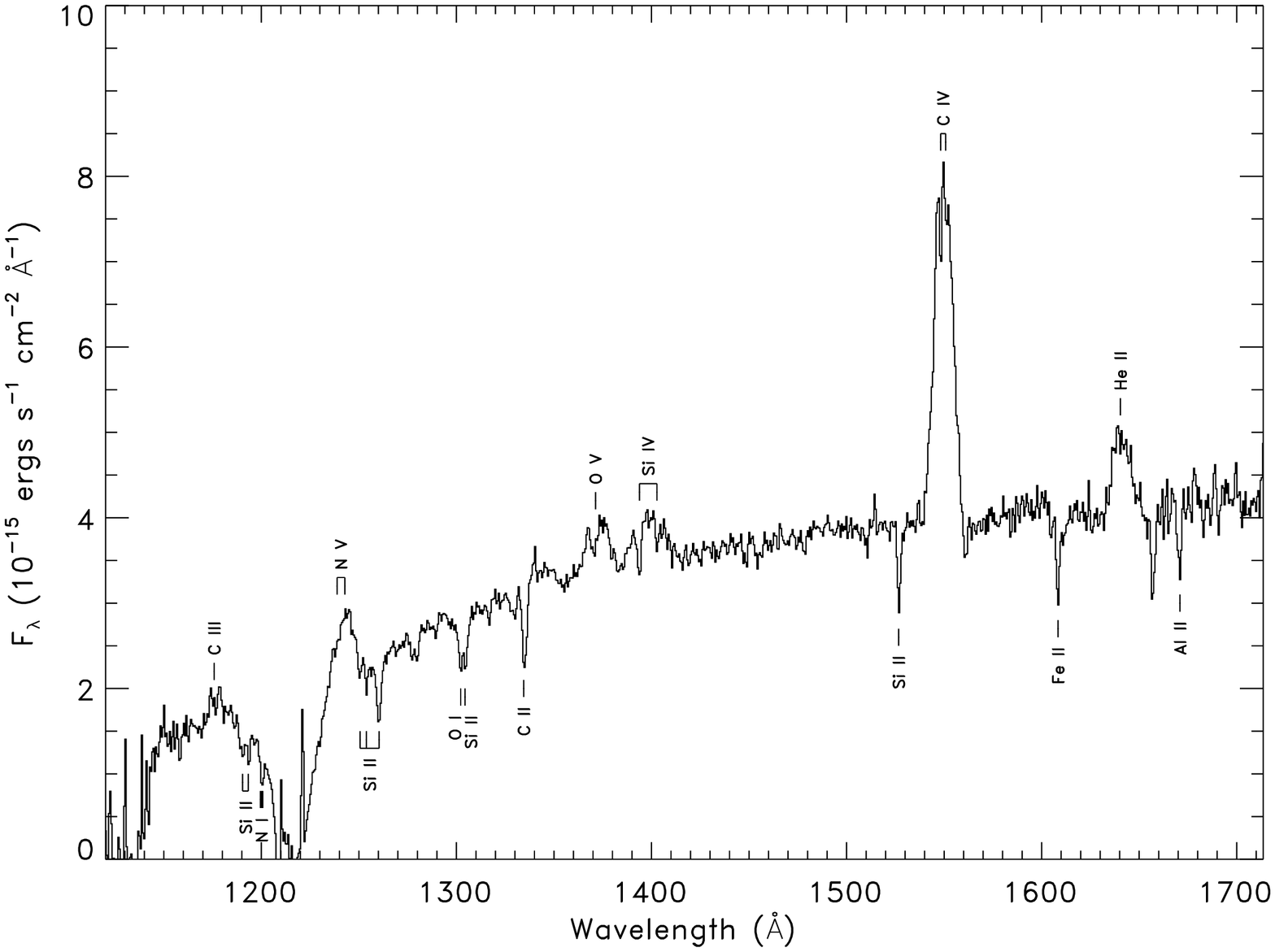}
\caption{XTE\,J1859+226.  Left: Broad band IR--UV SED at 
several epochs, corrected for interstellar reddening.
See text for details of the models applied.  Right: Average far
UV spectrum, uncorrected for reddening.  Identified lines are
marked.}
\end{figure}

The far-UV spectrum is rich in broad emission lines, with superposed
sharp interstellar absorption.  C\,{\sc IV}, He\,{\sc II}, N\,{\sc V},
O\,{\sc V} and Si\,{\sc IV} lines are definitely seen.  C\,{\sc IV}
shows emission to 4000\,km\,s$^{-1}$.  Ly$\alpha$ shows very broad
absorption, extending to 12000\,km\,s$^{-1}$

\vspace*{-4mm}
\section{Our New Target: XTE J1118+480}
\vspace*{-3mm}

XTE\,J1118+480 was discovered in March 2000 as a very weak X-ray
transient, although activity had started in January.  The optical
counterpart is bright ($V\sim13$), making the X-ray to
optical flux ratio exceptionally low.  The source is at high latitude
($+62^{\circ}$) and situated near the Lockman Hole, resulting in very
low absorption.  This combination of brightness and low absorption
makes it ideal for UV spectroscopy, and so we have been able use the
UV echelle modes of STIS. 

{\it HST} observations have been ongoing since April 8, with
coordinated {\it RXTE} and {\it EUVE} coverage.  We have found that
the IR--UV continuum is very flat (constant $F_{\nu}$;, Haswell, Hynes
\& King 2000, Hynes et al.\ 2000).  The UV spectrum shows emission
lines of He\,{\sc II}, N\,{\sc V} and Si\,{\sc IV}, but carbon and
oxygen are absent.  This likely indicates that the material being
transferred from the companion has been processed by the CNO cycle
within the interior, and hence that the envelope has been lost
(Haswell, Hynes \& King 2000).  The source exhibits remarkable rapid
variability on timescales of seconds at all wavelengths where such
variations can be detected.  This includes the UV where STIS TIMETAG
mode allowed resolution of variability up to 1\,Hz or higher.  A
strong correlation is seen between the X-ray and UV light curves, with
the UV lagging by 1--2\,s behind the X-rays (Haswell et al.\ 2000).

\vspace*{-4mm}
\section{Conclusions}
\vspace*{-3mm}

{\it HST} has now observed a number of BHXRTs in outburst.  Progress
has been made towards understanding these objects, but new questions
have been raised, both by observations and theoretical developments.
The wide wavelength coverage and reliable flux calibration of {\it
HST} make it an unrivalled tool for determining the IR--UV spectrum of
BHXRTs.  The diversity of SEDs observed so far, however, indicates
that more than one source of optical/UV emission may be present, and
careful modeling will be required to unambiguously separate these
contributions.  These include an accretion disk heated either by
viscosity or X-ray irradiation, the companion star and possible
synchrotron emission.  Exploitation of {\it HST}'s rapid spectroscopy
capability via echo-mapping provides additional clues to locate the
emission source within the binary, as demonstrated for GRO\,J1655--40
and potentially for XTE\,J1118+480.  These and other techniques allow
us to watch the evolution of the components of the binary through an
outburst, and hence improve our understanding of the outburst
mechanism and the nature of accretion onto black holes.

\vspace*{-4mm}
\subsection*{Acknowledgements}
\vspace*{-3mm}

The results presented derive from HST/FOS programs GO4377, GO6017 and
HST/STIS programs GO8245 and GO8647.  We are grateful to the director
for allowing us to bring forward the latter from Cycle 9 to observe
XTE\,J1118+480.  We would like to thank our co-investigators and
collaborators on this project, especially Keith Horne, Chris Shrader,
Wan Chen, Wei Cui, Mario Livio, Sylvain Chaty, Kieran O'Brien, Chris
Mauche and Andrew King, as well as many others whom space precludes.
Thanks also for technical support to Tony Roman, Kailash Sahu, Jeff
Hayes, Tony Keyes, Ed Smith, Michael Rosa and all at STScI.  Financial
support for this project has been provided by NASA grants NAG5-3311,
GO4377-02-92A and GO-6017-01-94A, and also by grant F/00-180/A from
the Leverhulme Trust and grant SC1/180/96/326/G from the Nuffield
Foundation.

\vspace*{-4mm}
\subsection*{References}
\vspace*{-3mm}

Cannizzo J.K., 1999, in Disk Instabilities in Close Binary Systems,
S. Mineshige and J. C. Wheeler (eds.), UAP, p177\\
Charles P.A., 1998, in Theory of Black Hole Accretion Disks,
M. A. Abramowicz and G. Bjornsson and J. E. Pringle (eds.), CUP, p1\\
Cheng F.H., Horne K., Panagia N., Shrader C.R., Gilmozzi R., 
Paresce F., Lund N., 1992, ApJ, 397, 664\\
Esin A.A., Lasota J.-P., Hynes R.I., 2000, A\&A, 354, 987\\
Haswell C.A., Hynes R.I., King A.R., 2000, IAUC 7407\\
Haswell C.A., Skillman D., Patterson J., Hynes R.I., Cui W., 2000,
IAUC 7427\\
Hynes R.I., Haswell C.A., Shrader C.R., Chen W., Horne K., Harlaftis
E.T., O'Brien K., Hellier C., Fender R.P., 1998a, MNRAS, 300, 64\\
Hynes R.I., O'Brien K., Horne K., Chen W., Haswell C.A., 1998b, 
MNRAS, 299, L37\\
Hynes R.I., Haswell C.A., 1999, MNRAS, 303, 101\\
Hynes R.I., Haswell C.A., A.J. Norton, S. Chaty, 1999, IAUC 7294\\
Hynes R.I., Mauche C.W., Haswell C.A., Shrader C.R., Cui W., Chaty S.,
2000, ApJL, accepted\\
Narayan R., Mahadevan R., Quataert E., 1998, in Theory of Black Hole 
Accretion Disks, M. A. Abramowicz and G. Bjornsson and J. E. Pringle 
(eds.), CUP, p148\\
Tanaka Y., Shibazaki N., 1996, ARA\&A, 34, 607\\
Uemura M., Kato T., Pavlenko E., Shugarov S., Mitskevich M., 1999,
IAUC 7303\\

\end{document}